\documentstyle[12pt]{article}

\input epsf

\begin{document}
\begin{titlepage}
\today          \hfill
\begin{center}
\hfill    OITS-704 \\

\vskip .05in

{\large \bf
Universal Extra Dimensions and $b \rightarrow s \gamma$
}
\footnote{This work is supported by DOE Grant DE-FG03-96ER40969.}
\vskip .15in
K. Agashe \footnote{email: agashe@neutrino.uoregon.edu},
N.G. Deshpande \footnote{email: desh@oregon.uoregon.edu},
G.-H. Wu \footnote{email: gwu@electron.uoregon.edu}

\vskip .1in
{\em
Institute of Theoretical Science \\
University
of Oregon \\
Eugene OR 97403-5203}
\end{center}

\vskip .05in

\begin{abstract}

We analyze the effect of 
flat universal extra dimensions (i.e., extra dimensions 
accessible to all SM fields) on the process $b \rightarrow s \gamma$.
With one Higgs doublet, the dominant
contribution at one-loop is from Kaluza-Klein (KK) states
of the charged would-be-Goldstone boson (WGB) and of the top
quark. The resulting constraint on the size of the extra
dimension is comparable to the constraint from
$T$ parameter. In two-Higgs-doublet model II, the contribution
of zero-mode and KK states of
physical charged Higgs can cancel the contribution from
WGB KK states. Therefore, in this model,
there is no constraint
on the size of the extra dimensions from
the process $b \rightarrow s \gamma$ and also
the constraint on the mass of
the charged Higgs from this process is weakened compared to 
$4D$.   
In two-Higgs-doublet model I, the contribution
of
the zero-mode and KK states
of physical charged Higgs and that of the KK states of WGB
are of the same sign. Thus, in this model and for small $\tan \beta$, the
constraint on the size of the extra dimensions is stronger than in 
one-Higgs-doublet model and also 
the constraint on 
the mass of the charged Higgs is stronger than in $4D$. 

\end{abstract}

\end{titlepage}

\newpage
\renewcommand{\thepage}{\arabic{page}}
\setcounter{page}{1}

The
motivations for studying theories with {\em flat}
extra dimensions of size (TeV)$^{-1}$ accessible
to (at least some of) the SM fields
are varied: 
SUSY breaking \cite{anto}, 
gauge coupling unification \cite{ddg},
generation of fermion
mass hierarchies \cite{as} and electroweak symmetry
breaking by a composite Higgs doublet \cite{ewsb}.
From the $4D$ point of view, these extra dimensions
take the form of Kaluza-Klein (KK) excitations of SM fields with masses
$\sim n / R$, where $R$ is a typical size of an extra dimension.
In a previous paper \cite{us}, we observed that the contribution
of these KK states
to the process $b \rightarrow s \gamma$ might give a
stringent constraint on $R^{-1}$.
In this paper, we will analyze 
in detail the effects of these KK states on the process 
$b \rightarrow s \gamma$ both in models with one and two Higgs
doublets.

In models with {\em only} SM gauge fields in the bulk, 
there are contributions to muon decay, atomic parity violation 
(APV) etc.~from
tree-level exchange of KK states of gauge bosons
\cite{graesser, nath}. Then,
precision electroweak measurements result in a strong constraint 
on the size of extra dimensions and, in turn, imply that the effect
on the process $b \rightarrow s \gamma$ is small.

To avoid these constraints, 
we will focus on models with {\em universal} extra dimensions, i.e.,
extra dimensions accessible to {\em all} the SM fields.
In this case,
due to conservation of extra dimensional momentum, there are 
{\em no} 
vertices with only one KK state, i.e., coupling of 
KK state of gauge boson to 
quarks and leptons always involves (at least one) 
{\em KK} mode of quark or lepton.
This, in turn, implies that there
is no tree-level contribution to weak decays of quarks and leptons, APV 
$e ^+ e^- \rightarrow 
\mu ^+ \mu ^-$ etc.~from exchange of KK states of gauge bosons 
\cite{hall, appel}. 
However, there is a 
constraint on $R^{-1}$
from {\em one-loop} 
contribution of KK states of (mainly) the top quark to the
$T$ parameter. For $m _t \ll R^{-1}$, this constraint is
roughly given by 
$\sum _n m_t^2 \Big/ \left( m_t^2 + \left( n / R \right)^2 \right) 
\stackrel{<}{\sim} 0.5 - 0.6$ (depending on the neutral Higgs mass)
\cite{appel}. For the case of one extra dimension, this
gives $R^{-1} \stackrel{>}{\sim} 300$ GeV. The KK excitations of quarks
appear as heavy stable quarks at hadron colliders and searches by the
CDF collaboration
also imply $R^{-1}\stackrel{>}{\sim} 300$ GeV for one extra dimension
\cite{appel}.

We begin with an analysis of $b \rightarrow s \gamma$
for the case of minimal SM with one Higgs doublet
in extra dimensions.
 
\section{One Higgs doublet}
The effective Hamiltonian for
$\Delta S = 1$ $B$ meson decays is
\begin{eqnarray}
{\cal H}_{\hbox{eff}} & = &
\frac{4 G_F}{\sqrt{2}} \; V_{tb} V_{ts}^{\ast}
\sum _{j=1}^{8} C_j (\mu) {\cal O}_j,
\end{eqnarray}
where the operator relevant
for the transition $b \rightarrow s \gamma$ is  
\begin{eqnarray}
{\cal O}_7 & = & \frac{e}{16 \pi^2} \; m_b \; \bar{s}_{L \alpha}
\sigma ^{\mu \nu} b_{R \alpha} F_{\mu \nu}. 
\end{eqnarray}
The coefficient of this operator from $W-t$ exchange in the SM is
\begin{eqnarray}
C^{W}_7 (m_W) & = & - \frac{1}{2} A \left( \frac{m_t^2}{m_W^2} \right),
\label{c7w}
\end{eqnarray}
where the loop function $A$ is given by
\begin{eqnarray}
A (x) & = & x \left[ \frac{ \frac{2}{3} x^2 +
\frac{5}{12} x - \frac{7}{12} }{ \left( x - 1 \right) ^3 } -
\frac{ \left( \frac{3}{2} x^2 - x \right) \ln x }{ \left( x - 1 \right) ^4 }
\right].
\end{eqnarray}
Of course, this includes the contribution from
the charged would-be-Goldstone boson (WGB) (i.e., longitudinal $W$).

With extra dimensions, there is a
one-loop contribution from KK states of $W$ (accompanied by KK states
of top quark, $t^{(n)}$), 
but as we show below, this is smaller than
that from KK states of charged WGB.
In the limit $m_W \ll R^{-1}$,
the KK states of $W$ get a mass $\sim n / R$ by ``eating'' 
the field corresponding to extra
polarization of $W$ in higher dimensions 
\footnote{This can be seen from KK decomposition of 
fields in the $5D$ gauge kinetic term (see,
for example, appendix C of 2nd reference
in \cite{ddg}).} 
-- this field 
is a scalar from the $4D$ point of view.
Thus, the coupling of {\em all}
components of $W^{(n)}$ to fermions is $g$, unlike the case of 
the zero-mode,
where the coupling of {\em longitudinal} $W$ to fermions is given by 
the Yukawa coupling
of Higgs to fermions.
Therefore, 
the contribution of $W^{(n)}$
to the coefficient of 
the dimension-$5$ operator $\bar{s} \sigma _{\mu \nu} b F^{\mu \nu}$
is
$\sim e \; m_b \; g^2 / \left( 16 \pi^2 \right)
m_t^2 \sum _n 1 / \left( n / R \right) ^4$,
where
the factor $m_t^2$ reflects GIM cancelation. In terms of 
the operator ${\cal O}_7$, the 
contribution of each KK state of $W$ to
$C_7$ is $\sim m_t^2 m_W^2 / \left( n / R \right) ^4$.
 
From the above discussion, it is clear that
the {\em KK} states of charged would-be-Goldstone boson 
(denoted by WGB$^{(n)}$) are physical
(unlike the {\em zero}-mode). The loop
contribution 
of WGB$^{(n)}$ with mass $n / R$ (and $t^{(n)}$ with mass
$\sqrt{ m_t^2 + \left( n / R \right)^2 }$)
is of the same form as that of physical charged Higgs in
2 Higgs doublet models 
\cite{bsgamma} with the appropriate modification of
masses and couplings of virtual particles in the loop integral 
\begin{eqnarray}
C^{\hbox{\scriptsize WGB}^{(n)}}_7 
\left( R^{-1} \right)
& \approx & \frac{m_t^2}{ m^2_t + \left( n / R \right) ^2 } \left[
B \left( \frac{ m^2_t + \left( n / R \right) ^2 }
{ \left( n / R \right) ^2 } \right)
- \frac{1}{6} A \left( \frac{ m^2_t + \left( n / R \right) ^2 }
{ \left( n / R \right) ^2 } 
\right) \right].
\label{c7WGB}
\end{eqnarray}
Here, 
the factor $m_t^2 / \left( m^2_t + \left( n / R \right) ^2 \right)$ 
accounts for
(a)
the coupling of WGB$^{(n)}$ to
$t^{(n)}$ which is 
$\lambda _t \sim m_t / v$, i.e.,
the same as that of WGB$^{(0)}$ 
(longitudinal $W$), and (b) the fact that
this contribution decouples in the limit of large KK mass --
the functions $A$ and $B$ (see below) in
the above expression approach a constant as $n / R$ becomes 
large.

The loop function $B$ is given by 
\cite{bsgamma}
\begin{eqnarray}
B (y) & = & \frac{y}{2} \left[ \frac{ \frac{5}{6} y - \frac{1}{2} }
{ \left( y - 1 \right) ^2 } - \frac{ \left( y - \frac{2}{3} \right) 
\ln y }{ \left( y - 1 \right) ^3 } \right].
\end{eqnarray}
It is clear that the
ratio of the contribution of $W^{(n)}$ and that of WGB$^{(n)}$
is $\sim \left( m_W R / n \right)^2 \stackrel{<}{\sim} O(1/10)$ since
$R^{-1} \stackrel{>}{\sim} 300$ GeV
(due to constraints from the $T$ parameter
and searches for heavy quarks). In what follows, we will neglect the
$W^{(n)}$ contribution.   

At NLO, the coefficient of the operator at the scale
$\mu \sim m_b$ is given by \cite{buras}
\begin{eqnarray}
C_7 \left( 
m_b \right) & \approx 0.698 \; C_7 \left( 
m_W \right) - 0.156 \; C_2 \left( 
m_W \right) + 0.086 \; C_8 \left( 
m_W \right).
\label{c7mb}
\end{eqnarray}
Here, $C_2$ is the coefficient of the operator
${\cal O}_2 = \left( \bar{c}_{L \alpha} \gamma ^{\mu}
b_{L \alpha} \right) \left( \bar{s}_{L \beta} \gamma _{\mu}
c_{L \beta} \right)$
and is 
approximately same as in the SM (i.e., $1$)
since the KK states of $W$ do not contribute to it at tree-level. $C_8$ 
is the
coefficient of the chromomagnetic operator
${\cal O}_8 = g_s / \left( 16 \pi^2 \right) \; m_b \; \bar{s}_{L \alpha}
\sigma ^{\mu \nu} T^a_{\alpha \beta} b_{R \beta} G^a_{\mu \nu}$.
In the SM, $C_8 \left( m_W \right) \approx -0.097$ \cite{buras}
due to the contribution of $W-t$ loop
(using $m_t \approx 174$ GeV).
The coefficient of this operator
also gets a loop contribution from KK states which is
of the same order as the contribution to $C_7$.
Since the coefficient of 
$C_8$ in Eq. (\ref{c7mb})
is small, 
we neglect the contribution of KK states to $C_8$.
The coefficient of ${\cal O}_7$ at the scale $m_W$ is given by the sum of 
the contributions of $W^{(0)}$ (Eq. (\ref{c7w}))
and that of WGB$^{(n)}$ (Eq. (\ref{c7WGB})) summed over $n$
\footnote{We neglect the RG scaling of ${\cal O}_7$ between 
$R^{-1}$ and $m_W$.}. 
Since $C_7^W \left( m_W \right) < 0$ and 
$C_7^{ \hbox{\scriptsize WGB}^{(n)} }
\left( R^{-1} \right) > 0$,  
we see that
contribution from 
WGB$^{(n)}$ interferes  
destructively with the $W$ contribution.

The SM prediction for 
$\Gamma \left( b \rightarrow s \gamma \right)
/ \Gamma \left( b \rightarrow c l \nu \right)$
has an uncertainty of about $10 \%$ and the 
experimental error is
about $15 \%$ (both are $1 \sigma$ errors)
\cite{kn}. The central values
of theory and experiment agree to within $ 1/2 \; \sigma$.
The semileptonic decay is not affected
by the KK states (at tree-level).
Combining theory and experiment $2 \sigma$ errors in quadrature, 
this means that the $95 \%$ CL
constraint on
the contribution of KK states is that it should not 
modify the SM prediction for $\Gamma \left( b \rightarrow s \gamma
\right)$
by more than $36 \%$.
Since $\Gamma \left( b \rightarrow s \gamma \right) \propto
\left[ C_7 \left( 
m_b \right) \right] ^2$, the constraint is
$\bigg| \left[ C_7 ^{ \hbox{total} } 
\left( 
m_b \right)
\right] ^2 / \left[ C_7^{\hbox{\small SM}} \left( 
m_b \right) \right]^2 - 1 \bigg| 
\stackrel{<}{\sim} 36 \%$. 
Using $m_t \approx 174$ GeV, we get 
$A \approx 0.39$ in Eq. (\ref{c7w}) and $C_7^{\hbox{\small SM}} 
\left( m_b \right)
\approx - 0.3$
\footnote{The NNLO corrections
for the SM prediction of the rate for
$b \rightarrow s \gamma$ are also known
and are about a few percent.}
from Eq. (\ref{c7mb}). 
Assuming $m_t \ll R^{-1}$, we get $B \approx 0.19$
and $A \approx 0.21$ in Eq. (\ref{c7WGB}). Then, using
Eq. (\ref{c7mb}) and the above criterion, we get the
constraint 
\begin{equation}
\sum _n m_t^2 \Big/ \left( m^2_t + \left( n / R \right)^2 \right) 
\stackrel{<}{\sim} 0.5
\end{equation}
which is comparable to that from the $T$ parameter.
For one extra dimension, performing the sum over KK states
with the exact expressions for $A$ and $B$ in
Eq. (\ref{c7WGB}), the constraint is
$R^{-1} \stackrel{>}{\sim} 280$ GeV
\footnote{We assume that the extra dimension denoted by
$y$ is compactified on
a circle of radius $R$. The various fields are chosen to be either
even or odd 
under the $Z_2$ symmetry, $y \rightarrow -y$ as in \cite{appel}.
Thus, the summation is over positive integers $n$.}.

Next, we consider models with two Higgs doublets.  

\section{Two-Higgs-doublet model II}
In this case, 
contribution from zero-mode physical charged Higgs interferes constructively
with the $W$ contribution
\cite{bsgamma}: 
\begin{eqnarray}
C^{H^{+ \; (0)}}_{7, II} \left( m_W \right) & 
\approx & - B \left( \frac{m_t^2}{m_H^2} \right) - \frac{1}{6}
\cot ^2 \beta \; A
\left( \frac{m_t^2}{m_H^2} \right),
\label{c7II0}
\end{eqnarray}
where $\tan \beta$ is the ratio of vev's of the two Higgs doublets.
In $4D$, 
this contribution gives a strong constraint on charged Higgs mass, 
$m_H \stackrel{>}{\sim} 500$ GeV. 

The combined
effect from KK states of physical charged Higgs and WGB is:
\begin{eqnarray}
C_{7, II}^{\left( \hbox{\scriptsize WGB}^{(n)} + H^{+ \; (n)}
\right)} \left( R^{-1} \right) & \approx & \frac{m_t^2}{ m^2_t + 
\left( n / R \right) ^2 } \left[
B \left( \frac{ m^2_t + \left( n / R \right) ^2 }
{ \left( n / R \right) ^2 } \right) 
- \frac{1}{6} A \left( \frac{ m^2_t + \left( n / R \right) ^2 }
{ \left( n / R \right) ^2 }
\right)
\right. -
\nonumber
\\ 
 & & \left.
B \left( \frac{ m^2_t + \left( n / R \right) ^2 }
{ m_H^2 + \left( n / R \right) ^2 } \right) -
\frac{1}{6} \cot ^2 \beta \;
A \left( \frac{ m^2_t + \left( n / R \right) ^2 }
{ m_H^2 + \left( n / R \right) ^2 } \right) \right],
\label{c7IIKK}
\end{eqnarray}
where the first line is from KK states of WGB (Eq. (\ref{c7WGB})) and
the second line is from KK states of physical charged Higgs (KK analog of
Eq. (\ref{c7II0})). 

Assuming $m_H \sim O(R^{-1})$ or larger, 
the combined effect of KK states is typically
destructive with respect to the $W$ contribution. This is because 
$B \left( \frac{ m^2_t + \left( n / R \right) ^2 }
{ m_H^2 + \left( n / R \right) ^2 } \right)
< B \left( \frac{ m^2_t + \left( n / R \right) ^2 }
{ \left( n / R \right) ^2 } \right)$
and the $A$ contribution is small such that $C_{7, II}^{\left( WGB^{(n)} + 
H^{+ \; (n)} \right) } > 0$
\footnote{In the limit $m_H \ll R^{-1}$, 
the $B$'s cancel in Eq. (\ref{c7IIKK})
so that the combined effect of
KK states is constructive (and small) due to the $A$'s.}.
Thus, the contribution of zero-mode physical charged Higgs
can cancel that of KK states
so that there is no constraint on $R^{-1}$. Also, this implies that
the constraint 
on $m_H$ is weakened in the presence of extra dimensions
of size $O \left( m_H^{-1} \right)$ or larger.

\begin{figure}
\centerline{\epsfxsize=0.55\textwidth \epsfbox{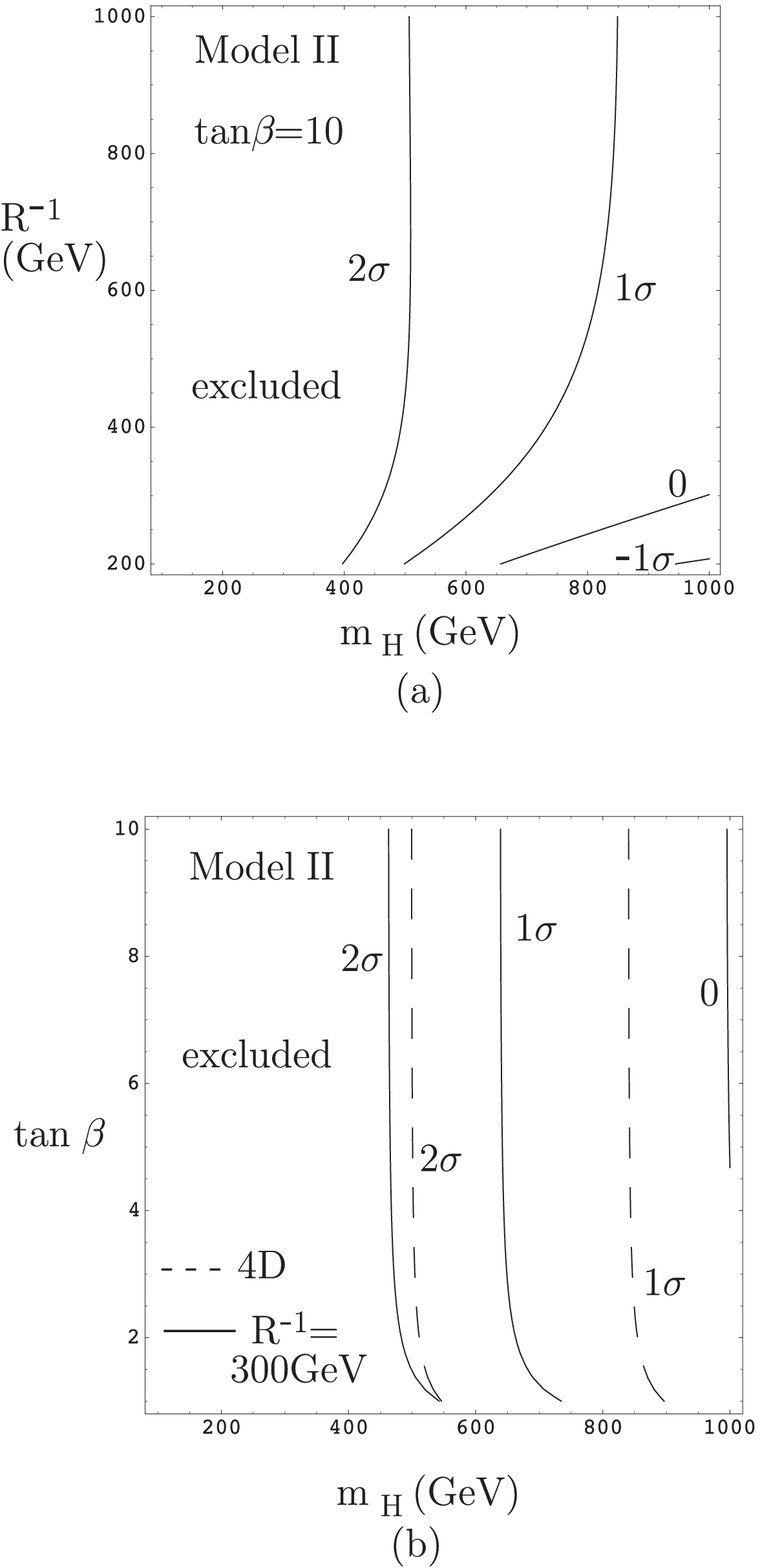}}
\caption{
The deviation 
of the rate of $b \rightarrow s \gamma$ from the SM prediction
in model II as a function of size of one extra dimension
($R^{-1}$) and charged Higgs
mass ($m_H$) for $\tan \beta = 10$ (figure (a)) and as a function
of $\tan \beta$ and $m_H$ for $R^{-1} = 300$ GeV
(figure (b)). In figure (b), the dashed lines are the result 
in $4D$. The $1 \sigma$ deviation corresponds to $ 18 \%$. 
}
\protect\label{bsgII}
\end{figure}

This can be seen in Fig. \ref{bsgII}
which shows the deviation 
in the rate for
$b \rightarrow s \gamma$ from the SM prediction for the case of one extra 
dimension. 
From Fig. \ref{bsgII}a,
we see that 
even for $R^{-1}$ as small as $200$ GeV \footnote{Of course, such
a small $R^{-1}$ might be ruled out
due to constraints from $T$ parameter and heavy quark searches.}, the
$95 \%$ CL constraint from $b \rightarrow s \gamma$ is satisfied
for a particular range of $m_H$. Of course, for $m_H \stackrel{>}{\sim}
1$ TeV, the effect of physical charged Higgs (both zero-mode and KK states)
becomes negligible so that we obtain the lower limit 
on $R^{-1}$ of about $300$ GeV
as in the one Higgs doublet case.
As seen from Fig. \ref{bsgII}b, 
the
$95 \%$ CL lower limit from $b \rightarrow s \gamma$ on $m_H$ is
about $500-550$ GeV (depending on $\tan \beta$) in $4D$.
We see that in
$5D$, the $95 \%$ CL lower limit on $m_H$ is reduced by about $40$ GeV
for $R^{-1} \sim 300$ GeV and the $1 \sigma$ limit on $m_H$ is reduced by
about $200$ GeV. 

\section{Two-Higgs-doublet model I}
In this case, the
contribution from zero-mode physical charged Higgs is destructive 
with respect to the $W$ contribution
\cite{bsgamma}:
\begin{eqnarray}
C^{H^{+ \; (0)}}_{7, I} \left( m_W \right) &
\approx & \cot ^2 \beta \left[ B \left( \frac{m_t^2}{m_H^2} \right)
- \frac{1}{6}
A
\left( \frac{m_t^2}{m_H^2} \right) \right].
\label{c7I0}
\end{eqnarray}
This contribution is negligible for large $\tan \beta$ and hence there is
no constraint on $m_H$ from $b \rightarrow s \gamma$ in $4D$. Of course, 
for small $\tan \beta$, this process does give a lower limit on
$m_H$: for $\tan \beta = 1$, the limit is about $350$ GeV.

The contribution from KK states
is also destructive with respect to the $W$ contribution:
\begin{eqnarray}
C_{7, I}^{\left( \hbox{\scriptsize WGB}^{(n)} + H^{+ \; (n)} \right)} 
\left( R^{-1} \right) & \approx & 
\frac{m_t^2}{ m^2_t +
\left( n / R \right) ^2 } \left[
B \left( \frac{ m^2_t + \left( n / R \right) ^2 }
{ \left( n / R \right) ^2 } \right)- 
\frac{1}{6} A \left( \frac{ m^2_t + \left( n / R \right) ^2 }
{ \left( n / R \right) ^2 }
\right) + \right.
\nonumber 
\\
 & & \left.
\cot ^2 \beta \left( B \left( \frac{ m^2_t + \left( n / R \right) ^2 }
{ m_H^2 + \left( n / R \right) ^2 } \right) -  
\frac{1}{6} A \left( \frac{ m^2_t + \left( n / R \right) ^2 }{
m_H^2 + \left( n / R \right) ^2 } \right) \right) \right], 
\end{eqnarray}
where the first line is from KK states of WGB (Eq. (\ref{c7WGB})) and
the second line is from KK states of physical charged Higgs (KK analog of
Eq. (\ref{c7I0})).
 
Thus, for small $\tan \beta$,
the constraint on $R^{-1}$ is stronger than with one
Higgs doublet and also the lower limit
on $m_H$ is larger with extra dimensions.

\begin{figure}
\vspace{0.3cm}
\centerline{\epsfxsize=0.55\textwidth \epsfbox{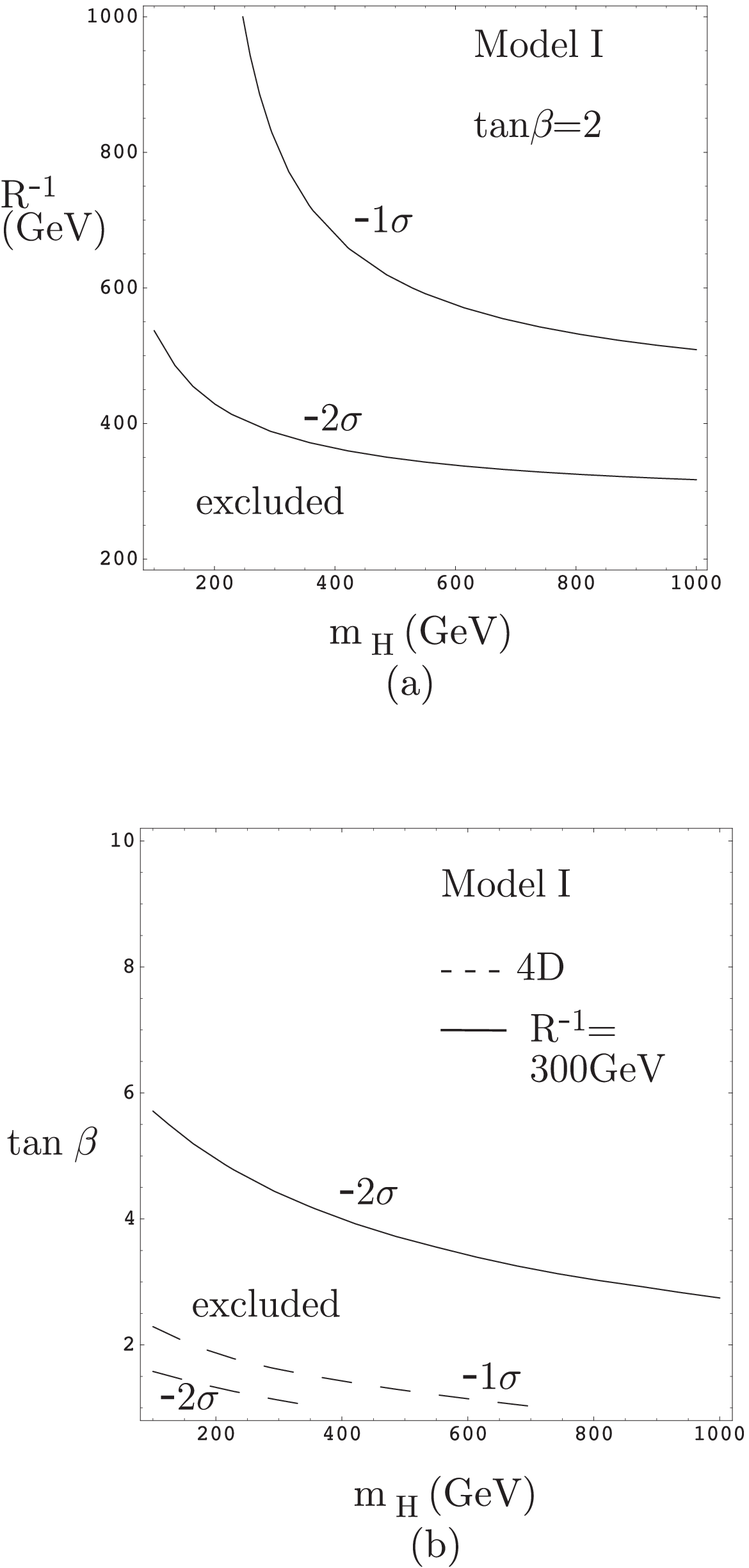}}
\caption{
The deviation 
of the rate of $b \rightarrow s \gamma$ from the SM prediction
in model I as a function of size of one extra dimension
($R^{-1}$) and charged Higgs
mass ($m_H$) for $\tan \beta = 2$ (figure (a)) and as a function
of $\tan \beta$ and $m_H$ for $R^{-1} = 300$ GeV
(figure (b)). In figure (b), the dashed lines are the result 
in $4D$.
The $1 \sigma$ deviation corresponds to $ 18 \%$.
}
\protect\label{bsgI}
\end{figure}

In Fig. \ref{bsgI}, we show the deviation from the SM prediction for
the rate of $b \rightarrow s \gamma$ for the case of one extra 
dimension.  
From Fig. \ref{bsgI}a, we see that 
for $\tan \beta = 2$ and $m_H
=100$ GeV, the lower limit on $R^{-1}$ is about
$550$ GeV
(as compared to about $300$ GeV in the one Higgs doublet case). 
Of course, for $m_H \stackrel{>}{\sim} 1$ TeV, the contribution of 
physical charged Higgs (both zero-mode and KK states)
is negligible and then the lower limit on $R^{-1}$ is the same as in
the one Higgs doublet case.
From Fig. \ref{bsgI}b, we see that
for $\tan \beta
=1$ and $R^{-1} = 300$ GeV, the limit on $m_H$ increases from about
$350$ GeV
in $4D$ to a value much larger than 
$1$ TeV.
As another example, for $\tan \beta = 4$, there is no constraint
on $m_H$ in $4D$, whereas for one extra dimension
of size $(300 \; \hbox{GeV})^{-1}$ there is a lower limit
on $m_H$ of about $400$ GeV. 
However, as in $4D$,
the effect of physical charged Higgs (both zero-mode and KK states) 
``decouples''
as $\tan \beta$ becomes larger and then we recover the one Higgs
doublet result for $b \rightarrow s \gamma$.

\section{Summary}
In this paper, we have studied the effect of 
universal extra dimensions on
the process $b \rightarrow s \gamma$. In the one Higgs doublet case, we showed
that the contribution of KK states of
charged would-be-Goldstone boson (WGB)
gives a constraint on the size of the 
extra dimensions which is comparable to that from the $T$ parameter. In
two-Higgs-doublet model
II, the contribution of physical charged Higgs 
(and its KK states) tends to cancel
the contribution of KK states of WGB so that there is no constraint on
the size of the extra dimensions and also the lower limit
on the charged Higgs mass is relaxed relative to
$4D$. 
In two-Higgs-doublet model I, the
contribution of physical charged Higgs 
(and its KK states) adds to the contribution of KK states
of WGB. Therefore,
for small $\tan \beta$,
the constraint on the size of extra dimensions becomes 
stronger than in the one-Higgs-doublet model and also the lower limit on
charged Higgs mass is larger than in $4D$.

\end{document}